\pdfoutput=1

\documentclass[reqno]{amsart}
\RequirePackage[l2tabu, orthodox]{nag} 

\usepackage{hyperref}

\usepackage{courier}
\usepackage{color}
\usepackage{amsmath,amsthm,amssymb}
\usepackage{cases} 
\usepackage[cal=boondox,scr=boondoxo]{mathalfa}

\usepackage{graphicx}
\usepackage[figurename=Fig.,labelsep=period]{caption}

\usepackage{textcomp} 
\usepackage{sistyle}
\usepackage{physics}
\usepackage{mathtools}
\mathtoolsset{showonlyrefs=false}

\theoremstyle{definition}
\newtheorem{remark}{Remark}

\usepackage[]{natbib}
\bibpunct[,]{(}{)}{,}{a}{}{,}
\newcommand{\citetapos}[1]{\citeauthor{#1}'s \citeyearpar{#1}}
\newcommand{\citetaposs}[1]{\citeauthor{#1}' \citeyearpar{#1}}

\begin{document}

\title[Total consensus under high reproductive variances]{
Total consensus under high reproductive-variance conditions}
\author[H.-S. Niwa]{H.-S. Niwa}
\date{}

\keywords{Pareto sampling;
 heterozygosity distribution;
 singularity;
 diversity paradox;
 marine species;
 conservation}

\begin{abstract}
Star-shaped branching patterns of genealogies are common in marine species. High-fecundity marine populations are characterized by low ratios of effective to actual population size, which reflect high variance in reproductive success among parents in mass spawns. When extreme reproduction events occur, offspring from very few parents dominate the population (whereby multiple mergers, or subsets of lineages with star-like trees, arise) and thus, the loss of genetic diversity is significant. Under high reproductive-variance conditions (assuming that reproduction occurs by sampling from the Pareto distribution), this paper explores the distribution of heterozygosity across generations. The result shows that zero heterozygosity is not achieved, implying that the populations may decline without evident loss of genetic variation. It is also found that there are singularities in the heterozygosity distributions. However, in the case of high reproductive variance, the locations of the singular points subtly deviate from those of the case where reproduction occurs by Wright-Fisher sampling.
\end{abstract}

\maketitle

\section{Introduction}
Genetic drift is changes in allele frequencies due to stochasticity (or variance) in the individuals' reproductive success within a population
\citep{Wright1931}.
Conventionally, the effect on genetic drift is accounted for by the effective population size.
Heterozygosity $H$, defined as the probability that any two randomly selected genes will be different,
is the most appropriate measure of genetic variability of a population
\citep{Nei1987}.
In a haploid population of effective size $N_{\mathrm{e}}$,
it follows that at equilibrium,
$H=2N_{\mathrm{e}}\mu/(2N_{\mathrm{e}}\mu+1)$ with neutral mutation rate $\mu$
\citep{Kimura-Crow64},
where there will be $N_{\mathrm{e}}\mu$ new mutants introduced per generation.
If $2N_{\mathrm{e}}\mu>1$, two or more alleles will usually be maintained.
If $N_{\mathrm{e}}$ is much less than $\mu^{-1}$ (so $H$ is low), all the genes in the population will usually be the descendants of a single mutant.

The classical model for genetic drift is the Wright-Fisher diffusion, which is appropriate when the reproductive variance is low
\citep{Kimura55-diffusion-eq},
i.e.
the family (or sibship) sizes are all very small compared to the population size $N$.
Genetic drift can be studied using the complementary or dual approach,
\citetapos{Kingman82-the-coalescent} coalescent,
where the probability of more than two lineages merging at each coalescent event is negligible, as $N\to\infty$.

Star-shaped genealogies have been reported in marine populations and explained by a post-glacial expansion following a Pleistocene population bottleneck
\citep[e.g.][]{Crandall-etal2012}.
Relying on the Wright-Fisher model,
these studies suggest that genetic drift
has been weak, so
allele frequencies have not changed rapidly in the post-expansion population.

Meanwhile, typical effective sizes of marine populations are in the hundreds or low thousands and
the $N_{\mathrm{e}}/N$ ratios may range from $10^{-5}$ to $10^{-2}$,
which are lower in larger populations
\citep{Hauser-Carvalho2008}.
Motivated by considering mass spawning species such as marine fishes, there has been interest in situations where, occasionally, a single family is of appreciable size compared with the population
\citep[i.e. sweepstakes reproductive success;][]{Hedgecock-Pudovkin2011}.
Low $N_{\mathrm{e}}/N$ ratios (i.e. high reproductive variances)
can alter allele frequencies significantly even for very large $N$.

Under Wright-Fisher sampling,
extreme reproduction events (whereby few individuals contribute most of the offspring to the next generation) will not occur.
Although the Kingman coalescent has proven robust to violations of most of its assumptions
\citep{Kingman2000},
it drastically fails to approximate the genealogies of species with high reproductive skew
\citep{Neher2013}.
Recent progress has been made on describing the genealogy of populations that occasionally have very large families
\citep[see][]{Tellier-Lemaire2014,Grant-ices2016,Montano2016},
where classical genetic drift is not adequate for describing random changes in allele frequencies.
When extreme reproduction events occur,
the temporal change in frequencies of alleles
can be described by a Wright-Fisher diffusion with jumps or a generalized Fleming-Viot process,
which is dual to the backward model, the $\Lambda$- or $\Xi$-coalescent
\citep{Donnelly-Kurtz99,Birkner2009-XiFV-recurrent-bottleneck,Der-etal2011}.
Jumps (or rapid alternations) in the allele frequency by genetic drift forward in time are equivalent to multiple mergers in its backward-time coalescent.

This paper examines the distribution of heterozygosity across generations.
The focus is to understand variation of reproductive success (i.e. variation of family sizes) within populations of species with high reproductive skew.
Recognizing the importance of maintaining low-frequency alleles
\citep{Franklin80},
it is necessary to investigate
whether reproductive skew will result in alleles drifting to fixation
(total consensus) within the populations,
which is still poorly understood under high reproductive-variance conditions (i.e. low $N_{\mathrm{e}}/N$ ratio conditions).
I explore the probability to have a total consensus with zero heterozygosity.

It has been reported that the observed histograms of heterozygosity have peaks or singularities
\citep[e.g.][]{Fuerst-etal1977},
and
comparisons between the measured distributions of $H$ and the predictions of the neutral theory were reviewed by \citet{Kimura83} as regards the validity of the neutral hypothesis of molecular evolution.
Here I show that,
although there are singularities in the $H$-distributions,
when reproduction is highly skewed,
the locations of the singular points subtly deviate from the expected norm \citep{Higgs-PRE95} for the infinitely-many-neutral-alleles Wright-Fisher model.

\section{Reproductive variance}
Assume a haploid population with a large number of reproducing individuals fixed at $N$ across generations.
Now consider Pareto sampling
\citep{Huillet2014,Huillet-Mohle2021,Niwa-arxiv10Feb2022}.
Let $Z_1,\ldots,Z_N$ be independent random variables
identically distributed according to the probability density
\begin{equation}\label{eqn:offspring-distr}
 f(z)=\alpha z^{-1-\alpha}
\end{equation}
with $\alpha>0$ and $z\geq 1$.
Upon normalizing the $Z_i$'s by their sum
\begin{equation*}
 R_N=\sum_{j=1}^N Z_j,
\end{equation*}
one defines the weight $W_i$ of the term $Z_i$ in the sum:
\begin{equation*}
 W_i=Z_i/R_N,
\end{equation*}
where $R_N/N^{1/\alpha}$, for large $N$, has a stable distribution called L{\'e}vy distribution
\citep[see][]{Bouchaud-Georges90}.
Each weight $W_i$ gives the probability of reproductive success of individual~$i$,
so the $i$-th family recruits a fraction $W_i$ of the offspring generation.
The weight is the normalized size of the family.
To put it another way, given the population at some generation, for each individual at the following generation, one chooses at random with probability $W_i$ one parent $i\in\{1,\ldots,N\}$.
Such a power-law offspring-number distribution
arises from type-III (exponential) survivorship with family-correlated survival
\citep{Reed-Hughes2002,Niwa-etal2017}.
Note that, when the $W_i$'s are identical for all $i$, the sampling procedure is equivalent to Wright-Fisher sampling.

Write $\rho(w)$ for the distribution of weight of families in the population,
such that $\rho(w)\dd{w}$ is defined as the expected number of families formed of size between $w$ and $w+\dd{w}$:
\begin{equation}\label{eqn:w-distr-def}
 \rho(w)=\mathrm{E}\qty[\sum_{i=1}^N \delta\qty(w-W_i)],
\end{equation}
which is the average of the empirical distribution for each generation
(i.e. each sample or realization),
where $\delta(\cdot)$ is the Dirac delta function.
The distribution $\rho(w)$ describes among-individual variation of reproductive success that exists in the population.
The probability of an individual (randomly sampled from the population) coming from a family of weight $w$ is given by $w\rho(w)$.
Since the reciprocal of the average weight of families gives the effective number of families (or reproducing lineages) in the population
\citep{Wright1931},
one has
\begin{equation*}
 N_{\mathrm{e}}=\qty[\int_0^1 w^2\rho(w)\dd{w}]^{-1},
\end{equation*}
which gives the proper time-scale for the asymptotic analysis of the ancestral process
\citep{Sagitov99}.
The ratio $N/N_{\mathrm{e}}$ represents the second moment (i.e. the variation) of reproductive success, which is also the effective family size in the population.

If the Pareto sampling is applied to the population, then for large $N$, Equation~\eqref{eqn:w-distr-def} reduces to
\begin{equation}\label{eqn:w-distr}
 \rho(w)=\frac{w^{-\alpha-1} (1-w)^{\alpha-1}}{N_{\mathrm{e}}{\,}\mathrm{Beta}(2-\alpha,\alpha)}
\end{equation}
with
\begin{equation*}
 N_{\mathrm{e}}=
 \left\{
  \begin{array}{ll}
   (1-\alpha)^{-1} & \quad\mbox{($0<\alpha<1$)}\\
   \ln N & \quad\mbox{($\alpha=1$)}\\
   \displaystyle{
   \frac{(\alpha/(\alpha-1))^{\alpha}N^{\alpha-1}}{\alpha\mathrm{Beta}(2-\alpha,\alpha)}} & \quad\mbox{($1<\alpha<2$)}
  \end{array}\right.
\end{equation*}
\citep{MezardPSTV84,Derrida-Flyvbjerg87,Derrida-PhysD97,Niwa-arxiv10Feb2022},
where $\mathrm{Beta}(a,b)=\mathrm{\Gamma}(a)\mathrm{\Gamma}(b)/\mathrm{\Gamma}(a+b)$ is the beta function.

When $1<\alpha<2$, although the distribution $\rho(w)$ as in Equation~\eqref{eqn:w-distr} diverges like $w^{-\alpha-1}$ for small $w$,
i.e. there are a large number of very small families, the effective family size is of order $N^{2-\alpha}$.
The population is then dominated by anomalously large families with size of order $N^{1/\alpha}$
\citep{Bouchaud-Georges90}.

In the case $\alpha<1$, Equation~\eqref{eqn:w-distr} implies that,
although families are mostly concentrated around $w=0$,
these small-$w$ families do not contribute to the total weight, as
\begin{equation*}
 \int_0^w w'\rho(w')\dd{w'}\propto w^{1-\alpha}.
\end{equation*}
So any particular one of them has an extremely small weight.
The probability $w\rho(w)$ is also peaked around $w=1$,
so there can be one large family in the interval $1/2<w\leq 1$.

\begin{remark}
Depending on the range of $\alpha$,
the $\rho(w)$ gives rise either to
a Poisson-Dirichlet$(\alpha,0){\;}$ $\Xi$-coalescent for $0<\alpha<1$,
or to a $\mathrm{Beta}(2-\alpha,\alpha)$ $\Lambda$-coalescent for $1\leq\alpha<2$.
When $0<\alpha<1$, the joint distribution of weights of families
(when ordering the weights decreasingly)
reduces to the two-parameter $\mbox{Poisson-Dirichlet}(\alpha,0)$ distribution
\citep{Pitman-Yor97}.
When $\alpha<2$,
$N_{\mathrm{e}} w^2\rho(w)$ is the $\mathrm{Beta}(2-\alpha,\alpha)$ density,
which reduces to the Dirac delta function (unit mass at zero) as $\alpha\to 2$.
When $\alpha\geq 2$ in Equation~\eqref{eqn:offspring-distr},
the variation of family sizes is low
(i.e. $N/N_{\mathrm{e}}$ is independent of $N$ for $\alpha>2$ or weakly (logarithmically) depends on $N$ for $\alpha=2$),
giving rise to the Kingman coalescent.
See \citet{Schweinsberg2003}, \citet{Huillet-Mohle2021}, and \citet{Niwa-arxiv10Feb2022}.
\end{remark}

\begin{remark}
The Kingman coalescent involves a linear time-scaling $N_{\mathrm{e}}\sim {N}$ if $\alpha>2$
(i.e. the ancestral tree height is proportional to $N$ generations),
or a time-scaling $N_{\mathrm{e}}=2N/\ln N$ for $\alpha=2$.
The symbol~$\sim$ denotes scaling, or asymptotic, equality up to a prefactor.
The $\mathrm{Beta}(2-\alpha,\alpha)$ $\Lambda$-coalescent involves a power-law time-scaling according to $N^{\alpha-1}$ for $1<\alpha<2$,
or a logarithmic time-scaling $\ln N$ for \citetapos{Bolthausen-Sznitman98} coalescent if $\alpha=1$,
where multiple mergers occur intermittently
(with intervals $\sim N_{\mathrm{e}}$ generations).
The $\mbox{Poisson-Dirichlet}(\alpha,0)$ $\Xi$-coalescent with $0<\alpha<1$
involves no time-scaling,
where multiple mergers occur,
independently of $N$,
at intervals of a few generations.
\end{remark}

\section{Total consensus}
This section examines how heterozygosity fluctuates in a single population (or equivalently, how it varies among loci).
Let $n_i$ be the number of copies of the $i$-th allele ($i=1,\ldots,K$) in the population of size $N$,
so that the frequency of the $i$-th allele is $x_i=n_i/N$.
Let $\phi(x)$ be the frequency spectrum, where $\phi(x)\dd{x}$ is defined as the equilibrium number of alleles expected in the frequency class $(x,x+\dd{x})$.
When taking a random sample from the population, the probability that the allele obtained has frequency between $x$ and $x+\dd{x}$ is $g(x)=x\phi(x)$.
Let $Y$ be the homozygosity,
\begin{equation*}
 Y=\sum_{i=1}^K x_i^2,
\end{equation*}
so that the heterozygosity is $H=1-Y$.
The probability distribution $\Pi_Y$ of homozygosity is defined as the average of the empirical distribution
\begin{equation*}
 \Pi_Y(y)=\mathrm{E}\qty[\delta\qty(\sum_{i=1}^K x_i^2-y)].
\end{equation*}
The average is taken over all possible partitions of the unit interval to $K$ pieces $\{x_1,\ldots,x_K\}$.

Under a diffusion approximation (where $\mu\ll 1$ and $N\gg 1$,
but the scaled mutation rate $\theta=2N_{\mathrm{e}}\mu$ is of order 1),
\citet{Kimura-Crow64} obtained
\begin{equation}\label{eqn:f-spectrum-kimura}
 \phi(x)=\theta x^{-1}(1-x)^{\theta-1}
\end{equation}
for the infinitely-many-neutral-alleles Wright-Fisher model
(in the limit $K\to\infty$ and $N\to\infty$ with constant $\theta$).
If $Y$ is close to 1,
the sum $Y$ is dominated by the very high-frequency alleles, and many of the low-frequency alleles in the population do not contribute much to $Y$.
Then, the homozygosity distribution is approximated as
\begin{equation*}
 \Pi_{Y\simeq 1}(y)\approx
  \int_{1-\varepsilon}^1 x\phi(x)\delta\qty(x^2-y)\dd{x} =
  2^{-\theta}\theta\qty(1-y)^{\theta-1}
\end{equation*}
(with small $\varepsilon>0$) from Equation~\eqref{eqn:f-spectrum-kimura},
implying that the distribution of the heterozygosity behaves like $H^{\theta-1}$ close to $H=0$.
Thus, in the Wright-Fisher model, there is a divergence as $H\to 0$ if $\theta<1$.
In other words,
if the scaled mutation rate $\theta$ is less than the critical value of 1,
no genetic variation (total consensus) in the population can be observed
\citep[see also][]{Donnelly-Kurtz96_Fleming-Viot,Durrett-CFP99}.
To say that the probability is concentrated near $H=0$ is to say that most of the time (high-frequency) alleles are near fixation.

Under high reproductive-variance conditions,
the value of $N_{\mathrm{e}}$ is expected to be very low
(expressed by the parameter $\theta<1$).
For example,
\citet{Niwa-etal2017} obtained $\alpha=1.24$ and $\theta=0.85$
for Japanese sardine (\textit{Sardinops melanostictus}) mitochondrial DNA cytochrome b sequences.
Nevertheless, the $\mathrm{Beta}(2-\alpha,\alpha)$ $\Lambda$-coalescent
predicts a much lower probability to have a total consensus with $H=0$
than the Kingman coalescent,
both with the same $N_{\mathrm{e}}$ given by $\theta<1$
\citep{Niwa-etal2017}.
See also \citet{Der-etal2011} and \citet{Montano2016}.
However, in yet another class of coalescents arising from the reproductive variation (Equation~\ref{eqn:w-distr}),
the $\Xi$-coalescent,
the computation of the probability of alleles drifting to fixation
is much more challenging.
In the following subsections,
I generate probability distributions of the heterozygosity by simulation of
the process of genetic drift
and neutral mutation
(which equivalently provides the interlocus variation of heterozygosity).
At each generation each individual (gene) experiences a mutation at rate $\mu=\theta/(2N_{\mathrm{e}})$,
where each new mutation creates an allele that has never before existed in the population.

\subsection{Wright-Fisher sampling}
\begin{figure}[htb]
 \centering
 \includegraphics[height=.5\textwidth,bb=0 0 300 301]{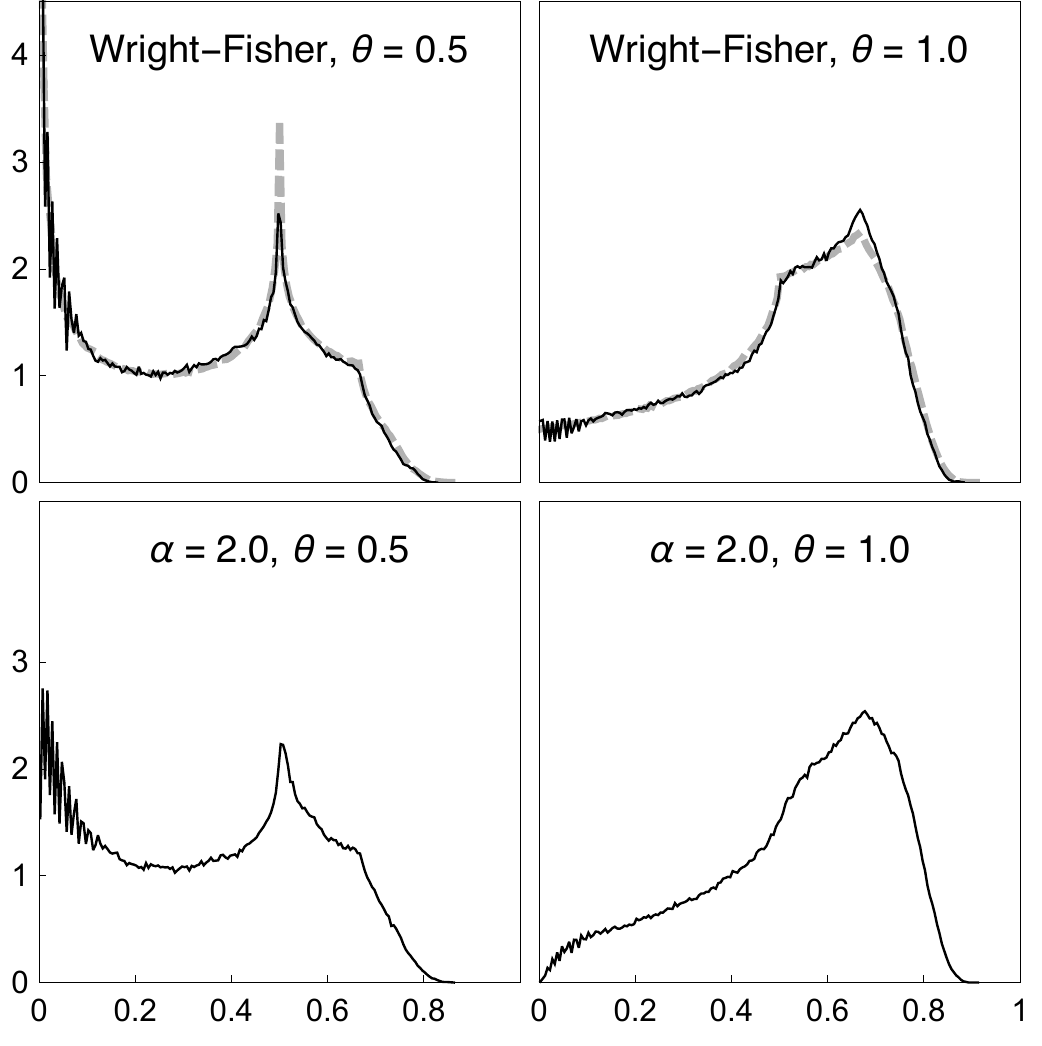}
 \caption{\small
Probability distribution of $H$ for a population with $N=10^3$.
The noisy curves were calculated by explicit simulation of the sampling models,
while the gray dashed curves were generated using $10^8$ iterations of the recursion (Equation~\ref{eqn:Derrida-Flyvbjerg-trick} in \S\ref{sect:singularity}) with the probability distribution $g(x)=x\phi(x)$ given in Equation~\eqref{eqn:f-spectrum-kimura}
\citep[see also][]{Derrida-Flyvbjerg87,Higgs-PRE95,Niwa-arxiv10Feb2022}
}\label{H-distr-a2-WF}
\end{figure}

I performed simulations of the random sampling and mutation (with two values of $\theta=0.5$ and $1.0$) using a population of $N=10^3$.
The reproduction model employed in the simulation is either Wright-Fisher sampling or Pareto sampling with $\alpha=2$.
Figure~\ref{H-distr-a2-WF} shows the histograms of the heterozygosity $H$ obtained after $10^6$ generations,
which gives the probability distribution of $H$ across many realizations of the process.
The numbers of hits on $H=0$
for $\theta=0.5$ and $1.0$ are, respectively, 28,693 and 885 in the case of Wright-Fisher sampling, and 1,288 and 1 in the case of $\mathrm{Pareto}(\alpha=2)$ sampling.

From \citetaposs{Ewens72} sampling formula with $\theta=0.5$ and $1.0$,
the probabilities to have a total consensus in a sample of 100 sequences are $8.87\times 10^{-2}$ and $0.01$, respectively.
Replicate samples (each of 100 sequences) from the Kingman coalescent with $\theta=0.5$ (resp. $1.0$) were generated with the program \texttt{ms}
\citep{Hudson02},
and 89,149 (resp. 10,085) of $10^6$ replications showed total consensus.

The probability distributions of $H$ have singularities (sharp changes in derivative) at all the values $1-1/\ell$ with $\ell=2,3,\ldots$, which become less pronounced with increasing $\ell$
\citep{Derrida-Flyvbjerg87}
and also for larger values of $\theta$ \citep{Higgs-PRE95}.
The pattern of the distribution shown in Figure~\ref{H-distr-a2-WF} (in the case of Wright-Fisher sampling or $\mathrm{Pareto}(\alpha=2)$ sampling, both with $\theta=0.5$) is similar to the observed histogram of heterozygosity over many loci in most species studied by \citet{Nei-etal76}, \citet{Fuerst-etal1977}, and \citet{Singh-Rhomberg87}.
Due to the non-self-averaging property of homozygosities,
i.e. the large interlocus variation in $H$
\citep{Stewart76,Higgs-PRE95},
fairly large numbers of loci would have to be examined in order to accurately characterize the heterozygosity of a population.

\subsection{Beta coalescent}
\begin{figure}[htb]
 \centering
 \includegraphics[height=.335\textwidth,bb=0 0 360 236]{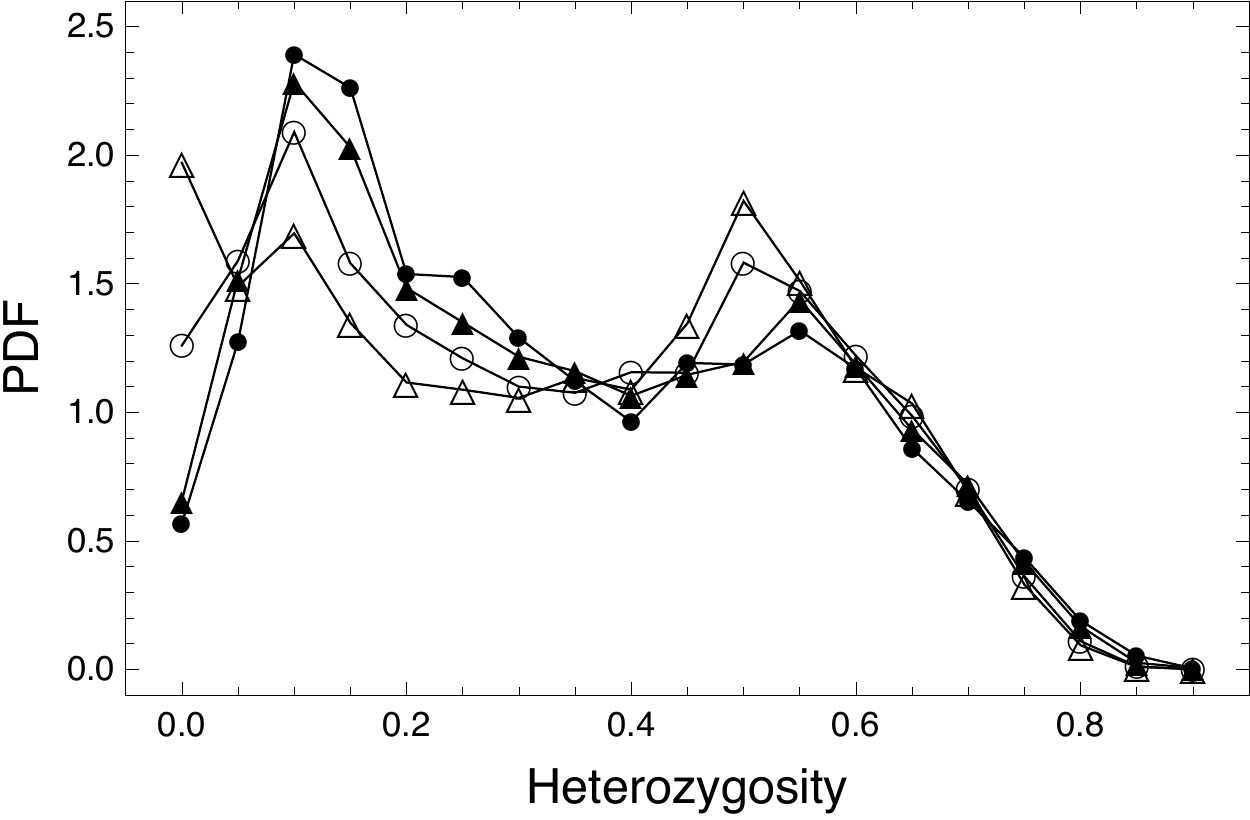}
 \caption{\small
Probability distribution of $H$ for a population with $\theta=0.5$.
The histograms are plotted for $\alpha=1.0$ (solid circles), $1.25$ (solid triangles), $1.5$ (open circles), and $1.75$ (open triangles)
}\label{H-distr-Beta-coal-t05}
\end{figure}

I simulated $10^4$ replications of 100 sequences
for the infinite-alleles model ($\theta=0.5$) under the $\mathrm{Beta}(2-\alpha,\alpha)$ coalescent with four values of $\alpha=1.0, 1.25, 1.5$, and $1.75$
through the time-reversed block counting process
\citep{BB08}.
Figure~\ref{H-distr-Beta-coal-t05} shows that
1.06\%, 0.96\%, 2.37\%, and 4.83\% of the samples have, respectively, the total consensus sequences.
When $1<\alpha<2$,
genetic variation can be retained for a longer period of time
in the populations with lower $N_{\mathrm{e}}/N$ ratios (i.e. higher reproductive variances with smaller $\alpha$).
There are singularities in the resulting $H$-distributions.
However, when $\alpha=1.0$ and $1.25$, the positions shift from the expected value $H=1/2$.
%

\subsection{Pareto sampling}
I further performed simulations of the random Pareto sampling and mutation ($\theta=0.5$) using a population of $N=10^3$,
and four values of $\alpha=0.9, 1.0, 1.25$, and $1.5$,
where the numbers of hits on $H=0$ are, respectively, 0, 0, 1, and 2 during the $10^6$ generations of the process.
A total consensus will hardly be reached among the population, and low-frequency alleles will usually be maintained.

\begin{figure}[htb]
 \centering
 \includegraphics[height=.5\textwidth,bb=0 0 300 301]{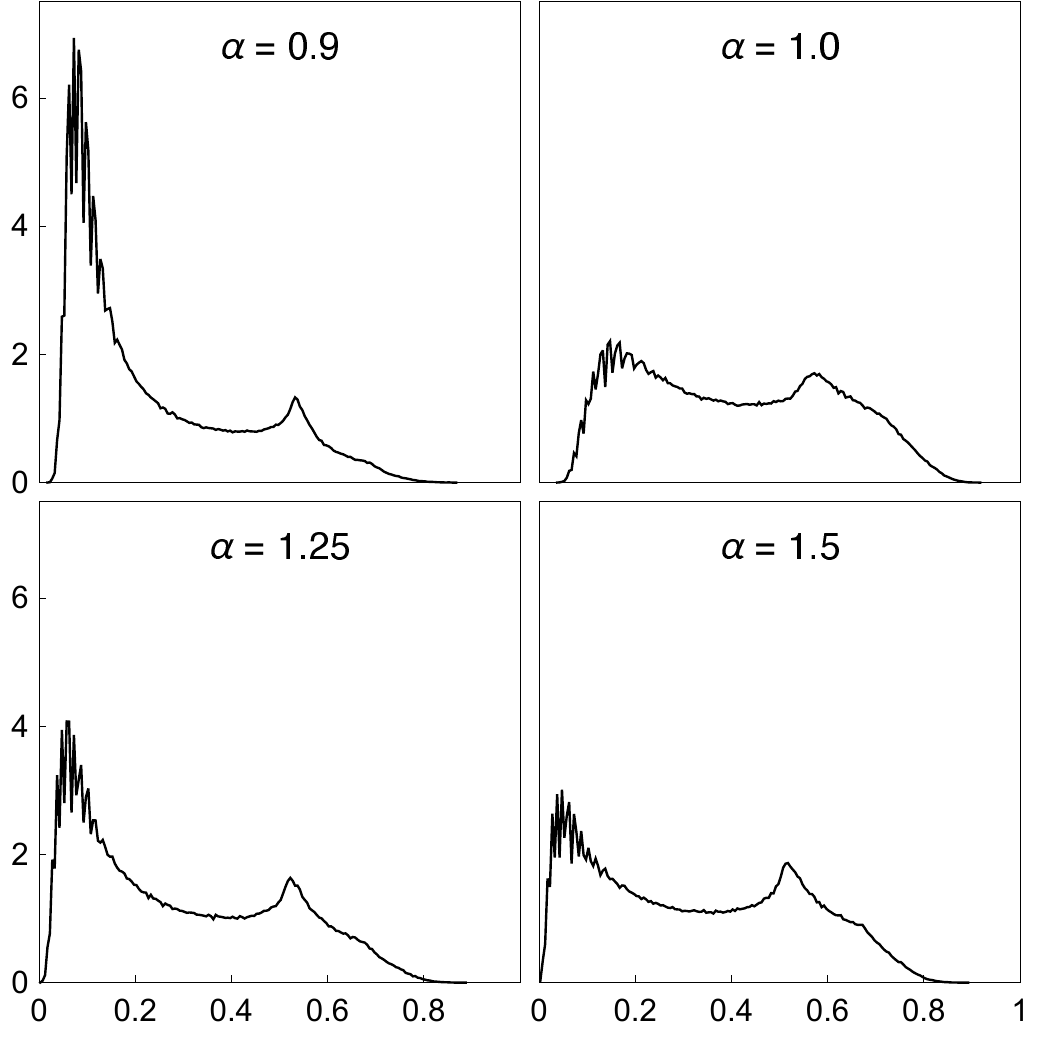}
 \caption{\small
Probability distribution of $H$ for a population with $N=10^3$ and $\theta=0.5$
}\label{H-distr-t05}
\end{figure}

Figure~\ref{H-distr-t05} shows the probability distribution of $H$ across realizations of the process.
The heterozygosity distributions are bimodal (M-shaped), and there are no divergences as $H\to 0$.
The distribution has a cutoff at the lower end (close to zero), denoted by $\hat{H}{\,}(\neq 0)$.
One then observes a peak in coalescence $\tau\approx\hat{H}/(2\mu)$ generations ago;
very few pairs (or groups) of lineages coalesce since then
\citep{Neher2013}.
Within the neutral Kingman coalescent framework, a distribution of this kind would be interpreted as a rapid population expansion starting around $\tau$ generations ago.
However, the size of the population did not change (with very low $N_{\mathrm{e}}$) in the simulation.

The simulation results show that,
while the $H$-distributions have singularities, the singular points for $\alpha<2$ shift from the locations
($H=1/2,2/3,\ldots$)
found for Wright-Fisher sampling (or Pareto sampling with $\alpha=2$).

\section{Singularities in the heterozygosity distribution}\label{sect:singularity}
Now, following \citet{Derrida-Flyvbjerg87},
let us discuss the (shifting of) singularities,
assuming that the density function $\Pi_Y$ of homozygosity diverges at $Y=1-\hat{H}$ with exponent $\beta_1<0$, i.e.
with $\Pi^{(\mathrm{sing})}$ denoting the dominant part of the singularity in $\Pi_Y$,
\begin{equation}\label{eqn:singular-1}
 \Pi^{(\mathrm{sing})}\qty(Y\simeq (1-\hat{H}))\propto\abs{Y-(1-\hat{H})}^{\beta_1}.
\end{equation}
Let us begin with $K{\,}(\gg 1)$ different alleles in the population of size $N{\,}(\gg 1)$ and homozygosity $Y_K$.
The frequency of the $K$-th allele is $x_K$,
so that the number of copies of the $K$-th allele in the population is $Nx_K$.
The $N(1-x_K)$ remaining individuals of the other $K-1$ alleles in the population should have the same statistical properties.
Therefore, one can write the recursion relation
\begin{equation}\label{eqn:Derrida-Flyvbjerg-trick}
 Y_K=x_K^2+(1-x_K)^2 Y_{K-1},
\end{equation}
where $Y_{K-1}$ corresponds to the homozygosity in the remaining population.
So, $Y_K$ and $Y_{K-1}$ have the same limiting distribution $\Pi_Y$ for $K\to\infty$.
The distribution $\Pi_Y$ is nothing but the histogram of the $Y_K$, where the sequence $Y_K$ is constructed by the random process given by Equation~\eqref{eqn:Derrida-Flyvbjerg-trick}.
That is,
one has a new $Y$ value given by $\tilde{Y}=x^2+(1-x)^2 Y$,
where $x$ is randomly chosen according to a given probability distribution $g(x)$.
Iterating this formula requires one random variable $x$ for each new $Y$ value created
(cf. also Figure~\ref{H-distr-a2-WF}).
From Equation~\eqref{eqn:Derrida-Flyvbjerg-trick} one sees the $\Pi_Y$ obeys the integral equation
\begin{align}\label{eqn:integral-eq}
 \Pi_Y(y) &=\int_0^1 g(x)\dd{x}\int_0^1\Pi_Y(y')\dd{y'}
 \delta\qty(y-x^2-\qty(1-x)^2 y')\nonumber\\
 &=\int_0^1\frac{g(x)\dd{x}}{\qty(1-x)^2}\Pi_Y\qty(\frac{y-x^2}{(1-x)^2}).
\end{align}
Now consider the function
\begin{equation*}
 Y'(x)=\frac{Y-x^2}{\qty(1-x)^2}.
\end{equation*}
For $Y$ close to $1/\ell-\hat{H}/\ell^2$ for $\ell=2,3,\ldots$, i.e.
$Y=1/\ell-\hat{H}/\ell^2+\varepsilon$
with small $\varepsilon>0$,
one has
\begin{equation*}
 Y'(x) = \frac1{\ell-1}-\frac{\hat{H}}{(\ell-1)^2}+
 \qty(\frac{\ell}{\ell-1})^2\varepsilon-\qty(\frac{\ell}{\ell-1})^3\qty(x-\frac1\ell)^2
 +\order{\qty(x-\frac1\ell)^3},
\end{equation*}
where $x_{-}<x<x_{+}$ with
\begin{equation*}
 x_{\pm}=\frac1\ell \pm\qty(\frac{\ell-1}{\ell}\varepsilon)^{1/2}
\end{equation*}
(double-sign corresponds).
Consequently the integral in Equation~\eqref{eqn:integral-eq} picks up a contribution from
$\Pi_Y\qty(Y\simeq\frac1{\ell-1}-\frac{\hat{H}}{(\ell-1)^2})$.
Thus, for the singularity
\begin{equation*}
 \Pi^{(\mathrm{sing})}\qty(Y\simeq\frac1{\ell-1}-\frac{\hat{H}}{(\ell-1)^2})\propto
  \qty(Y-\qty(\frac1{\ell-1}-\frac{\hat{H}}{(\ell-1)^2}))^{\beta_{\ell-1}}
\end{equation*}
with exponent $\beta_{\ell-1}$,
one gets
\begin{align*}
 \Pi^{(\mathrm{sing})}\qty(\frac1\ell-\frac{\hat{H}}{\ell^2}+\varepsilon)
 &\propto
 \int_{x_{-}}^{x_{+}}\dd{x}
 \qty[\qty(\frac{\ell}{\ell-1})^2\varepsilon-\qty(\frac{\ell}{\ell-1})^3\qty(x-\frac1\ell)^2]^{\beta_{\ell-1}}\\
 &\propto\varepsilon^{\beta_{\ell-1}+1/2},
\end{align*}
where $g(1/\ell)<\infty$ is assumed.
Since $Y'=1-\hat{H}$ for $Y=1/2-{\hat{H}}/{4}$,
the singularity of $\Pi_Y$ at $Y=1-\hat{H}$ gives rise to a singularity at $Y=1/2-{\hat{H}}/{4}$.
This singularity may be viewed as `inherited' from the singularity at $Y=1-\hat{H}$.
In general, $Y'=1/\ell-\hat{H}/\ell^2$ for $Y=1/(\ell+1)-\hat{H}/(\ell+1)^2$ with $\ell=1,2,\ldots$, yielding
\begin{equation}\label{eqn:singular-ell}
 \Pi^{(\mathrm{sing})}\qty(Y\simeq\qty(\frac1\ell-\frac{\hat{H}}{\ell^2}))\propto
  \qty(Y-\qty(\frac1\ell-\frac{\hat{H}}{\ell^2}))^{(\ell-1)/2+\beta_1}
\end{equation}
with $\beta_1$ as in Equation~\eqref{eqn:singular-1}.
An integer exponent in Equation~\eqref{eqn:singular-ell} (for $\beta_1=-1/2$) signals a logarithmic singularity.

I did not prove the existence of the singularity of Equation~\eqref{eqn:singular-1} at $Y=1-\hat{H}$.
The simulation results imply that
the distribution for allele frequencies has a cutoff
at some high frequency $\hat{x}<1$, i.e.
for $x$ close to $\hat{x}$,
\begin{equation}\label{eqn:singular-g}
 g(x)\propto (1-x/\hat{x})^{\beta_1}
\end{equation}
with $\beta_1$ being the same exponent as in Equation~\eqref{eqn:singular-1}.
Now, consider the case $\hat{x}\simeq 1$,
then $\hat{x}^2\simeq 1-\hat{H}$.
For $Y_\ast=1-\hat{H}-\varepsilon$
(with small $\varepsilon>0$), from Equation~\eqref{eqn:integral-eq} one has
\begin{equation*}
 \Pi_Y(Y_\ast)\approx\frac{g(Y_{\ast}^{1/2})}{(1-Y_{\ast}^{1/2})^2}
 \int_{x_\ast}^{Y_{\ast}^{1/2}}\dd{x}\Pi_Y\qty(\frac{Y_\ast-x^2}{(1-x)^2}),
\end{equation*}
where $x_\ast$ satisfies
\begin{equation*}
 \frac{Y_\ast-x_{\ast}^2}{(1-x_\ast)^2} = \hat{x}^2,
\end{equation*}
yielding
\begin{equation*}
 x_{\ast}^2\approx Y_\ast-\frac12 \qty(\hat{H}+\varepsilon)^2.
\end{equation*}
Because $\Pi_Y$ is normalized, one sees that
\begin{equation*}
 \Pi_Y(Y_\ast)\approx\frac{g(Y_{\ast}^{1/2})}{2Y_{\ast}^{1/2}}
  \propto\varepsilon^{\beta_1}
\end{equation*}
from the singularity of $g(x)$ at $\hat{x}$ given in Equation~\eqref{eqn:singular-g}.

In summary,
the values $H_\ell=1-1/\ell+\hat{H}/\ell^2$ (with $\ell=1,2,\ldots$) are singular points of the probability distribution of heterozygosity.
The locations of the singular points depend on $\alpha{\,}(<2)$ as in Equation~\eqref{eqn:offspring-distr}, through the $\alpha$ dependence of $\hat{H}{\,}(>0)$.
Note that $\hat{H}\to 0$ as $\alpha\to 2$,
and
in the case $\alpha\geq 2$, the locations of the singular points are identical to those found for Wright-Fisher sampling.

\section{Conclusions}
The skewed distribution of reproductive success is widely observed among marine species.
When reproduction is highly skewed, the value of $N_{\mathrm{e}}$
is expected to be very low.
The potential consequences highlight the importance of maintaining evolutionary potential.
Maintaining low-frequency alleles is important to the long-term maintenance of populations under changing environmental conditions.
They are not presently beneficial, although, in the future, they may be of use.
Understanding genetic drift in species of conservation concern is essential for designing better conservation strategies.
The application of classical population genetics theory would, however, be inappropriate for high-fecundity marine species.

This paper is concerned with the effect of reproductive skew on genetic drift.
Reproduction occurs by $\mathrm{Pareto}(\alpha)$ sampling.
If $\alpha\geq 2$, the reproductive variance is low,
so the allele frequency follows classical genetic drift,
i.e.
drift removes allelic variation from the population continuously at a rate inversely proportional to population size
(the genetic diversity or the $N_{\mathrm{e}}$ of a population is proportional to its size $N$).
If $\alpha<2$,
classical genetic drift is not even approximately real in the broad distribution of family sizes.
Jumps in the allele frequencies occur intermittently with time-intervals of order $N_{\mathrm{e}}$ generations.

$N_{\mathrm{e}}$ is a measure of how many individuals contribute to the next generation.
In populations with low $N_{\mathrm{e}}/N$ ratios (i.e. high reproductive variances),
a tiny minority of adults will wind up being the parents of the vast majority of the next generation.
We have seen that, even for small $N_{\mathrm{e}}$ (less than the critical size expressed by $\theta=1$),
some genetic variation may be observed under the $\mathrm{Beta}(2-\alpha,\alpha)$ $\Lambda$-coalescent process.
Further, I have provided the first analysis of this topic for $\Xi$-coalescent processes.
The probability distribution of heterozygosity generated by Pareto sampling for reproduction (in the case $\alpha<1$ and $\theta<1$)
has a rather sharp cutoff at the lower end,
so that a total consensus with $H=0$ will never be reached among the population.

Think of the scenario where few lineages substantially contribute to the next generation.
This scenario expresses single lineage's rapid expansion,
from which it follows that no scaled time would pass and no coalescence occur.
In classical terms, rapid growth stops genetic drift,
resulting in star-shaped branching patterns of (local) genealogies.
However, this scenario does not imply an expansion of the population size, which can remain constant.
Rather, an actual population may decline substantially without evident loss of genetic variation.
Some studies have documented such a diversity paradox: overexploited and depleted marine fish populations with temporal stability in genetic diversity
\citep[see][and references therein]{Niwa-etal2017}.
For example, \citet{Ruzzante-etal2001} observed genetic stability for Newfoundland populations of Atlantic cod (\textit{Gadus morhua}) even though the population today is only 1\% of its former size.
The maintenance of genetic diversity in a collapsed stock is not predicted by classical population genetic theory.
Under $\Lambda$- and $\Xi$-coalescents,
rare alleles can persist for a longer time than under Kingman coalescent,
implying a reduced probability of loss (resp. fixation) for very low-frequency (resp. high-frequency) alleles.
By contrast, when reproductive variance and $N_{\mathrm{e}}$ are low (with $\theta<1$),
alleles at low frequencies are more likely to be lost by drift.

In the infinite-alleles model the number $K$ of distinct alleles actually present in the population fluctuates.
The probability distribution $\Pi_Y$ of homozygosity has a different functional form in each interval $1/K$ to $1/(K+1)$,
so that the $\Pi_Y$ has singularities
\citep{Derrida-Flyvbjerg87,Higgs-PRE95}.
I have shown that,
although there are singularities in the heterozygosity distribution,
when reproduction is highly skewed (and $\theta<1$),
the locations of the singular points subtly deviate from the expected norm for the infinitely-many-neutral-alleles Wright-Fisher model.
The fact that the total consensus with $H=0$ is hard to reach
explains the shifting of singularities of the distribution.

The genetic diversity of a population is shaped by its recent demographic history.
The shallow genealogy of the star-shaped coalescent might be a signature of rapid population growth following a bottleneck
(as occurs, for example, during the last glaciation).
Detecting multiple mergers in genetic data is important for understanding which forces have shaped the diversity (e.g. an excess of low-frequency alleles) of a population: post-glacial expansion vs. reproductive skew (with weak vs. strong genetic drift).
The compatibility of a sample with the standard (neutral) Kingman coalescent is typically assessed using \citetapos{Tajima89a} $D$-statistic.
One approach to identifying multiple mergers
is to use the site-frequency spectrum as a summary statistic
\citep[see][]{Neher2013,Tellier-Lemaire2014}.
However, it is not trivial to determine which coalescent model is applied to account for observed genetic data
\citep[see][]{Niwa-etal2016}.
The shifting of singularities of the heterozygosity distribution is thought to be a signature of high reproductive skew in a population.
When examining the interlocus variation of heterozygosity in the sample,
by looking at the locations of singular points of the heterozygosity distribution,
the subtle deviations (compared with the Wright-Fisher model)
could help distinguish between the Kingman and multiple-merger coalescents.
It would be interesting to know whether these shifting of singularities can be observed in marine species.
Also it would be interesting to know what information about the process of genetic drift (reproductive stochasticity) and neutral mutation is contained in the knowledge of the shifting of singularities of the heterozygosity distribution.

\bibliographystyle{apalike2}
\bibliography{bib-niwa-genetics}

\end{document}